\documentclass{article}

\usepackage{arxiv}

\usepackage[utf8]{inputenc} 
\usepackage[T1]{fontenc}    
\usepackage{hyperref}       
\usepackage{url}            
\usepackage{booktabs}       
\usepackage{amsfonts}       
\usepackage{nicefrac}       
\usepackage{microtype}      
\usepackage{lipsum}
\usepackage{graphicx}
\usepackage{caption}
\usepackage{subcaption}
\usepackage{multirow}
\usepackage{amsthm}
\usepackage{amsmath}
\graphicspath{ {./images/} }

\title{Identifying early-warning indicators of tipping points in networked systems against sequential attacks}

\author{
 Utkarsh Gangwal \\
  Civil Engineering\\
  Indian Institute of Technology Gandhinagar\\
  Gandhinagar, 382355 \\
  \texttt{utkarsh.gangwal@iitgn.ac.in} \\
   \And
 Udit Bhatia* \\
  Civil Engineering\\
  Indian Institute of Technology Gandhinagar\\
  Gandhinagar, 382355 \\
  \texttt{bhatia.u@iitgn.ac.in} \\
  \And
 Mayank Singh \\
  Computer Science and Engineering\\
  Indian Institute of Technology Gandhinagar\\
  Gandhinagar, 382355 \\
  \texttt{} \\
  \And
 Pradumn Kumar Pandey \\
  Computer Science and Engineering\\
  Indian Institute of Technology Roorkee\\
   \\
  \texttt{} \\
  \And
 Deepak Kamboj \\
  \\
  Indian Institute of Engineering Science and Technology, Shibpur\\
   \\
  \texttt{} \\
  \And
 Samrat Chatterjee \\
  \\
  Pacific Northwest National Laboratory\\
   \\
  \texttt{} \\
}

\begin{document}
\maketitle
\begin{abstract}
Network structures in a wide array of systems such as social networks, transportation, power and water distribution infrastructures, and biological and ecological systems can exhibit critical thresholds or tipping points beyond which there are disproportionate losses in the system functionality. There is growing concern over tipping points and failure tolerance of such systems as tipping points can lead to an abrupt loss of intended functionality and possibly non-recoverable states. While attack tolerance of networked systems has been intensively studied for the disruptions originating from a single point of failure, there have been instances where real-world systems are subject to simultaneous or sudden onset of concurrent disruption at multiple locations. Using open-source data from the United States Airspace Airport network and Indian Railways Network, and random networks as prototype class of systems, we study their responses to  synthetic attack strategies of varying sizes.  For both types of networks, we observe the presence of warning regions, which serve as a precursor to the tipping point. 
Further, we observe the statistically significant relationships between network robustness and size of simultaneous distribution, which generalizes to the networks with different topological attributes for random failures and targeted attacks. We show that our approach can determine the entire robustness characteristics of networks of disparate architecture subject to disruptions of varying sizes. Our approach can serve as a paradigm to understand the tipping point in real-world systems, and the principle can be extended to other disciplines to address critical issues of risk management and resilience.
\end{abstract}

\keywords{Tipping points $|$ Warning regions$|$ Large-scale Impacts $|$ Indian Railways $|$ US Airspace}

\section*{Introduction}
There is growing concern over the onset of the sudden collapse in the networked systems ranging from natural systems including ecosystems to built systems including the Internet, multi-modal transportation networks, power grids, and water distribution systems \cite{pocock2012robustness,luke2010power,ganin2017resilience,yazdani2012applying,buldyrev2010catastrophic,wang2014robustness,jackson2001historical,svendsen2007connectivity,yletyinen2019understanding}. Examples of such collapse are accelerated extinction of species in ecosystems \cite{dakos2019ecosystem,lever2014sudden}, nation-wide blackouts in the power-grid as a consequence of localized failures that further percolated into interdependent systems including communication and transportation networks \cite{buldyrev2010catastrophic}, and of regional and local transportation systems because of accidents, weather events, and congestion \cite{ganin2017resilience}. When networks are subject to such events, a fraction of components (nodes, edges, or combination thereof) may disappear, resulting in the loss of intended functionality.

Investigating attack tolerance for both single and interdependent networked systems has been a central question given growing concern over nation-state attacks on built systems, and extinction of species as a consequence of environmental degradation \cite{gao2011robustness,buldyrev2010catastrophic,dong2013robustness, svendsen2007connectivity}.  Researchers have investigated the system dynamics near a “point of no return” or tipping points using a set of non-linear dynamical equations to the stability of these systems \cite{jiang2018predicting,gao2016universal,morone2019k,jiang2019harnessing}. While for specific systems, such dynamics are relatively well understood, encapsulating the complex interactions that shape up present real-world complex systems in the form of an analytical framework is a non-trivial task. Moreover, the complex systems with a large number of interacting components can pose a challenge of prohibitively high dimensionality. Thus, researchers in the past have examined the robustness characteristics of networked systems by linking it to the structure of the underlying networks \cite{albert2000error,barabasi2016network,bhatia2015network,bollobas2004robustness}. 
While attack tolerance of networked systems has been intensively studied for the disruptions originating from a single point of failure, from both topological and dynamical perspectives, there have been instances where real-world systems are subject to simultaneous or sudden onsets concurrent of disruption at multiple locations. To understand the tolerance of systems toward sequential attacks,  it is critical to analyse system vulnerability to  repeated attacks of varying sizes, and proactively identify their tipping points to multiple disruptions. Moreover, once the tipping point sets in, stakeholders and infrastructure managers may not have  opportunities to course-correct for preventing catastrophic system loss. Hence, identifying the warning regions preceding the tipping points can enable stakeholders to plan for graceful degradation before systems enter into irrecoverable or prolonged damage states \cite{steinberg2011baton,gariel2008graceful,scheffer2012anticipating}. 

To address these gaps, we develop a method to identify the onset of warning point and region as well as tipping point while accounting for multiple attacks and recurring faults occurring in the networks, both real-world and synthetic, of disparate sizes. We seek to address whether warning regions can be identified for the networks before tipping points are reached, when networks are subject to sequential attacks. Using a combination of graphical and numerical techniques, we find that while tipping points for networks of different types and sizes exhibit considerable variability, the onset of the warning region exhibits a consistent pattern for all the networks. Further, we note that network robustness (quantified as the area under robustness curve), and batch size (number of nodes removed at each instance) exhibit linear relationships on logarithmic scales and hence offering methods to predict robustness characteristics for the systems under recurring attacks.

\section*{Data}
We demonstrate  applicability of the proposed approach on both real-world networks and synthetic networks (Table S1). Here we model Indian Railways Network (IRN), and US National Airspace System Airport Network (USNASAN) as undirected networks using open-source data, which consist of 809 and 1261 nodes, respectively. For IRN and USNASAN, nodes represent the stations and the airports, respectively. A pair of stations (airports) is considered to be connected if there is at least one direct train (flight) between them (See SI Table 1). 

We note that while the choice of these two real-world networks is guided by data availability \cite{bhatia2015network,clark2018resilience}, the methods can generalize to other systems that can be represented as networks. Since the structure of underlying networks plays an important role in determining system’s ability to survive targeted and random attacks \cite{molloy1995critical, bollobas2004robustness}, we use Erd\H{o}s R{\'e}nyi model to simulate random networks (hereinafter referred to as ER), and  Barab{\'a}si Albert model  to generate scale-free networks (hereinafter referred to as BA) \cite{barabasi2003scale} with the number of nodes varying from 1,000 to 25,000. To draw a direct comparison with the real-world network (IRN in this case), we keep the network density, $D$ (defined as the proportion of possible links to maximum possible links among the nodes) for synthetic network similar to that of IRN ($D_{IRN} = 0.008$). The four networks, used here, also follow different degree distributions that most real-world networks are known to exhibit \cite{albert2000topology,pagani2013power}. For example, IRN,  USNASAN, and BA networks follow power-law with degree distribution, $p_k$ of the form:

\begin{equation}
    p_k  = Ck^{-\gamma}
\end{equation}
where \textit{C} is the normalization constant and \textit{$\gamma$} is the degree exponent. 

On the other hand, for ER networks, $p_k$ follows binomial distribution of the form:
\begin{equation}
    p_k  = {n \choose k}p^k(1-p)^{n-k}
\end{equation}
 where \textit{n} is the number of nodes and $p$ is the probability of link between a pair of nodes (See SI: Fig. 1).  
 
\begin{figure*}[htb]
\centering
\includegraphics[scale = 0.8]{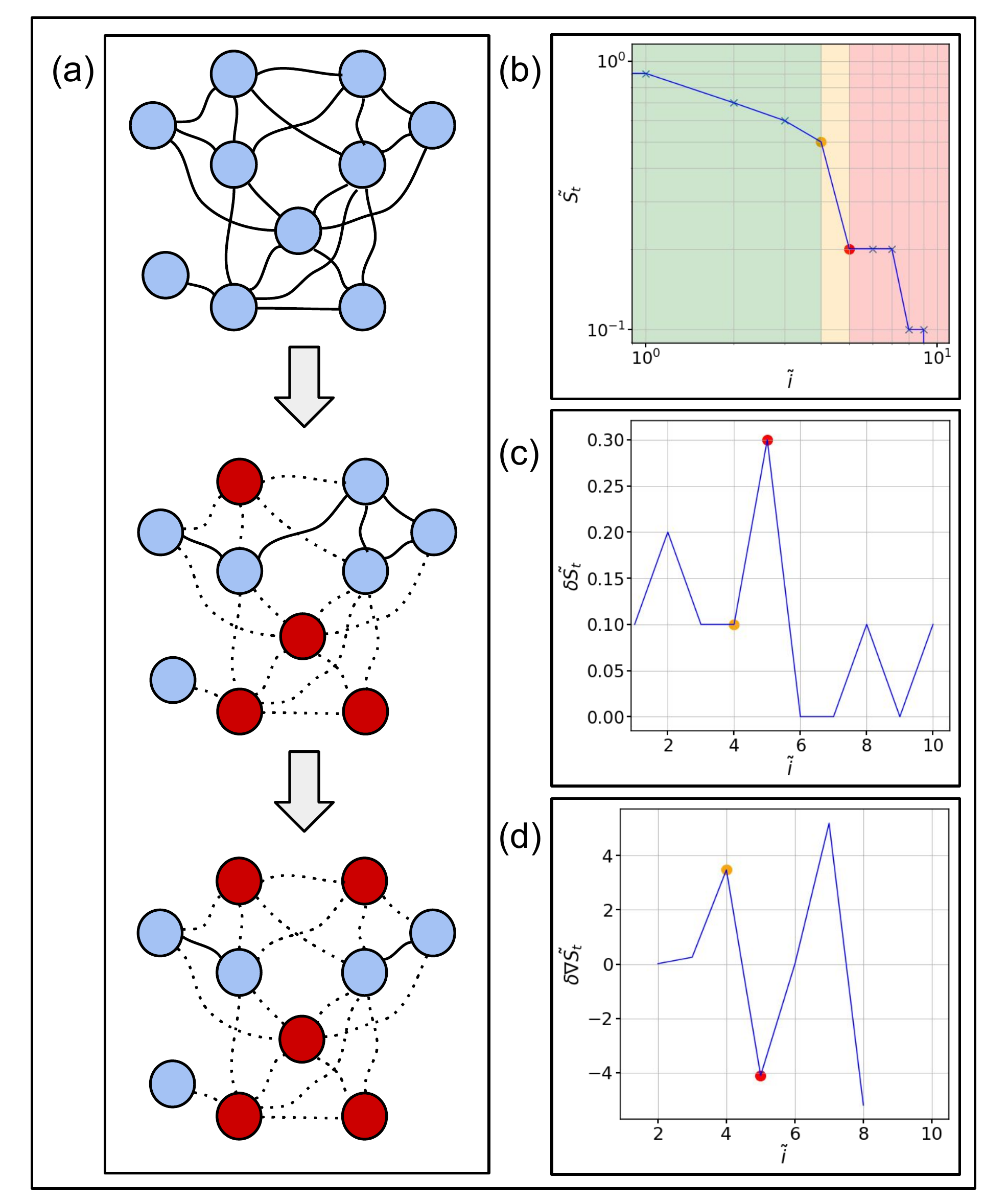}
\caption{\textbf{Warning region and tipping point} Illustrative example depicting the warning region and the tipping point for a network containing ten nodes with arbitrary degree distribution when perturbed with batch size, $B=1$.  \textbf{(a)} Represents the three distinct stages of the network as perturbation progresses. In the first stage, the network is in its original form and is fully functional (marks the starting point of the green region shown in \textbf{b}. As the perturbation progresses, deviation in the slope of $\Tilde{S_t}$ ($\delta\nabla{{\Tilde{S_t}}}$) begin to increase which is identified as warning point (shown in orange) in \textbf{d}. At warning point, the network still retains 50\% of its state of critical functionality. The onset of warning region (at $\Tilde{i}$=4) coincides with increased fluctuations in $\delta\nabla{{\Tilde{S_t}}}$ that continues even beyond tipping point.}
\label{fig:toy}
\end{figure*}

 Besides, we use a simplified illustrative network with ten nodes of arbitrary degree distribution to describe the terminologies related to the warning region and tipping points identified in this research (Fig. \ref{fig:toy}).

\section*{State of Critical Functionality}

To measure attack tolerance of the networked systems, including the Internet \cite{albert2000topology}, transportation systems \cite{bhatia2015network}, communications, and power distribution systems \cite{buldyrev2010catastrophic}, the measure of connectivity in terms of probability of finding a node in the largest connected cluster is often used. Here, we use a similar measure of the state of critical functionality, $\Tilde{S_t}$, which is defined as the ratio of the size of largest connected cluster at time  $t$ to its size before perturbation. We note that in the context of dynamical systems, the functionality is often measured in the form of network activities \cite{gao2016universal}. However, in the absence of known operational dynamics,  we adopt a proxy measure  ${\Tilde{S_t}}$ that provides extent of fragmentation of the largest connected cluster under various perturbation scenarios. Hence, $\Tilde{S_t}=1$ would represent a fully functioning network, whereas $\Tilde{S_t}=0$ represents the total loss of functionality.

\section*{Simulating perturbations}
The central question that this research attempts to address is: how networked systems respond to simultaneous disruptions occurring sequentially until the network loses its functionality? To model different attack scenarios, we consider the combination of random failures and targeted attacks. In each model, we synthetically disrupt remove a batch of nodes of size $B$ with $1 \leq B \leq n. $ The four disruption models are: 

\textbf{Random attack model (RAM):}  Randomly choose and remove $B$ nodes at each time-step. However, the selection is not straightforward due to stochastic nature. At each time-step, we randomly sample $K$ sets of $B$ nodes. For each set, we compute $\Tilde{S_t}$ and choose the set that yields the lowest $\Tilde{S_t}$ value (or worst possible scenario from given sampling during random perturbations).
  
\textbf{Targeted attack model (TAM-I):}
 At each time-step, top-$B$ nodes with high degrees are removed, and then the degrees of the remaining nodes are recomputed.

\textbf{Targeted attack model (TAM-II):}
 The probability of selection of a node depends on its current degree, i.e., higher the degree, the higher is the probability of being attacked. At each time-step, $B$ nodes are removed. Degrees are recalculated after each batch removal. 

\textbf{Targeted attack model (TAM-III):}
At each time-step, top-$B$ nodes with high degrees are removed first, followed by nodes with lower degrees. Unlike TAM-I, degrees are not recomputed.

\section*{Results}

\begin{figure*}[htb]
\centering
\includegraphics[scale = 0.8]{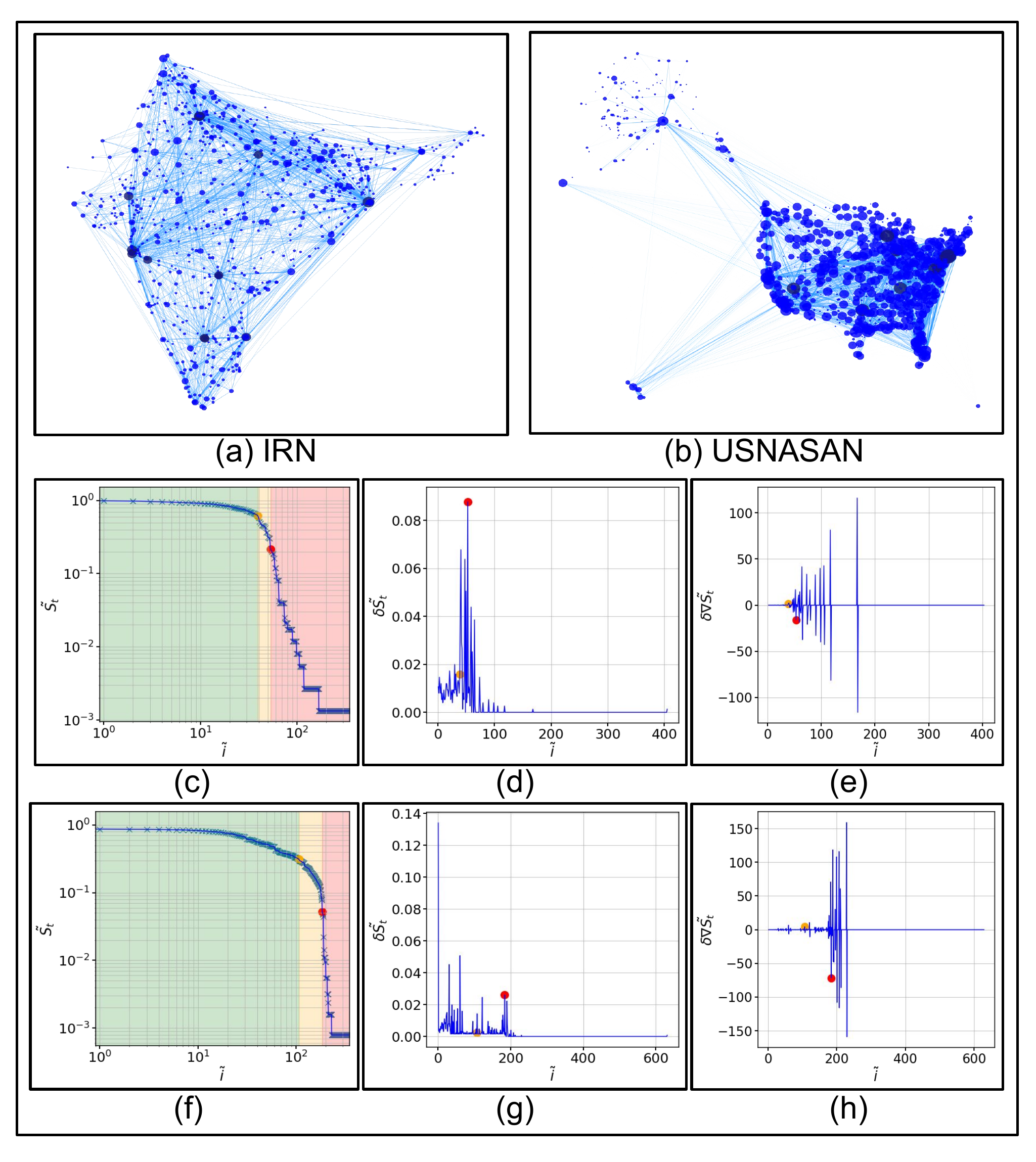}
\caption{\textbf{Delineating functionality regions in real-world networks} (a)Network representation of Indian Railways Network (IRN) and  \textbf{(b)} The U.S National Airspace System Airport Network (USNASAN). While IRN has 809 nodes and 2,342 edges, USNASAN has 1,261 nodes and 47,489 edges, depicting an edge density $\sim$7 times in comparison to IRN. (c) and (f) show the state of critical functionality ($\Tilde{S_t}$) vs. time instance ($\Tilde{i}$) for respective networks when these are perturbed using Targeted Attack Model (TAM-I) with batch size, $B$ = 2. While green region marks the region of high functionality, yellow and red regions mark the warning and tipping regions, respectively. (d) For IRN,while onset of tipping point is evident from $\delta{\Tilde{S_t}}$, (g) $\delta{\Tilde{S_t}}$ for USNASAN exhibit multiple spikes after the onset of warning point.(e) and (h) $\delta\nabla{{\Tilde{S_t}}}$ for  both IRN and USNASAN exhibit increased fluctuations in warning region. }
\label{fig:2}
\end{figure*}

In this research, we seek to address that when networked systems are subjected to sequential attacks of size $B$, whether warning regions can be identified before the point of sudden collapse or tipping point.  In agreement with the case of gradual removal of nodes reported in literature \cite{reka2000internet}, we notice that as the fraction of removed nodes increases, $\Tilde{S_t}$ decreases non-linearly exhibiting a threshold-like behavior with $\Tilde{S_t}$ reaching to zero. In addition to disproportionate loss in functionality, we  notice a large deviation in the size of largest connected component under all perturbation scenarios ($\delta{\Tilde{S_t}}$).  This point marked by the largest deviation is identified as \textit{tipping point}. Moreover, we note that before tipping point is reached under various perturbation scenarios, deviation in the slope of $\Tilde{S_t}$ ($\delta\nabla{{\Tilde{S_t}}}$) begins to increase. We measure the deviation in the slope of $\Tilde{S_t}$ on log-log scale and the  point after which the fluctuations in   $\delta\nabla{{\Tilde{S_t}}}$  begin to increase is referred to as \textit{warning point}. Despite the large fluctuations, for all networks considered in this study, $\Tilde{S_t}$ $>> 0$ at the warning point. Hence, the onset of the large fluctuations can serve as a precursor to the onset of sudden collapse. We identify the region between the warning point and the tipping point as  \textit{warning region}. 
We investigate  whether  two distinct real-world networks, IRN  and USNASAN (Fig. \ref{fig:2}(a) and \ref{fig:2}(b)),  exhibit warning and tipping points when subject to synthetic sequential disruptions of varying size $B$. For each network, we consider $B$ ranging from 2 to 98 with an increment of 2, resulting in 49 robustness curves.
While we identify the onset of warning regions using the graphical method (SI: Fig. 2), we recognize the tipping point as a point with the largest magnitude of $\delta{\Tilde{S_t}}$ following the warning point. Fig. \ref{fig:2}(c) and (f) show the robustness response of the two networks for $B=2$ with the warning region shown in yellow color. We note that the network density of USNASAN is 7.5 times higher than that of IRN. The greater density with a comparable number of nodes would translate to more redundancy, therefore  greater network robustness \cite{borgatti2006robustness,xie2019eradicating}. In the case of sequential disruptions based on synthetic failure schemes, we observe that warning and tipping points in  USNASAN occur later than in IRN. Also, the width of the warning region in USNASAN is greater compared to that of IRN. Moreover, the point of maximum deviation appears distinctly in the largest connected component after onset of the warning point, which is not the case in USNASAN  due to its high robustness whereby we observe multiple data spikes. We consider the spike with the largest magnitude as tipping point (Fig.~\ref{fig:2}(d) and~\ref{fig:2}(g)). 

\begin{figure*}[htb]
\centering
\includegraphics[scale = 0.8]{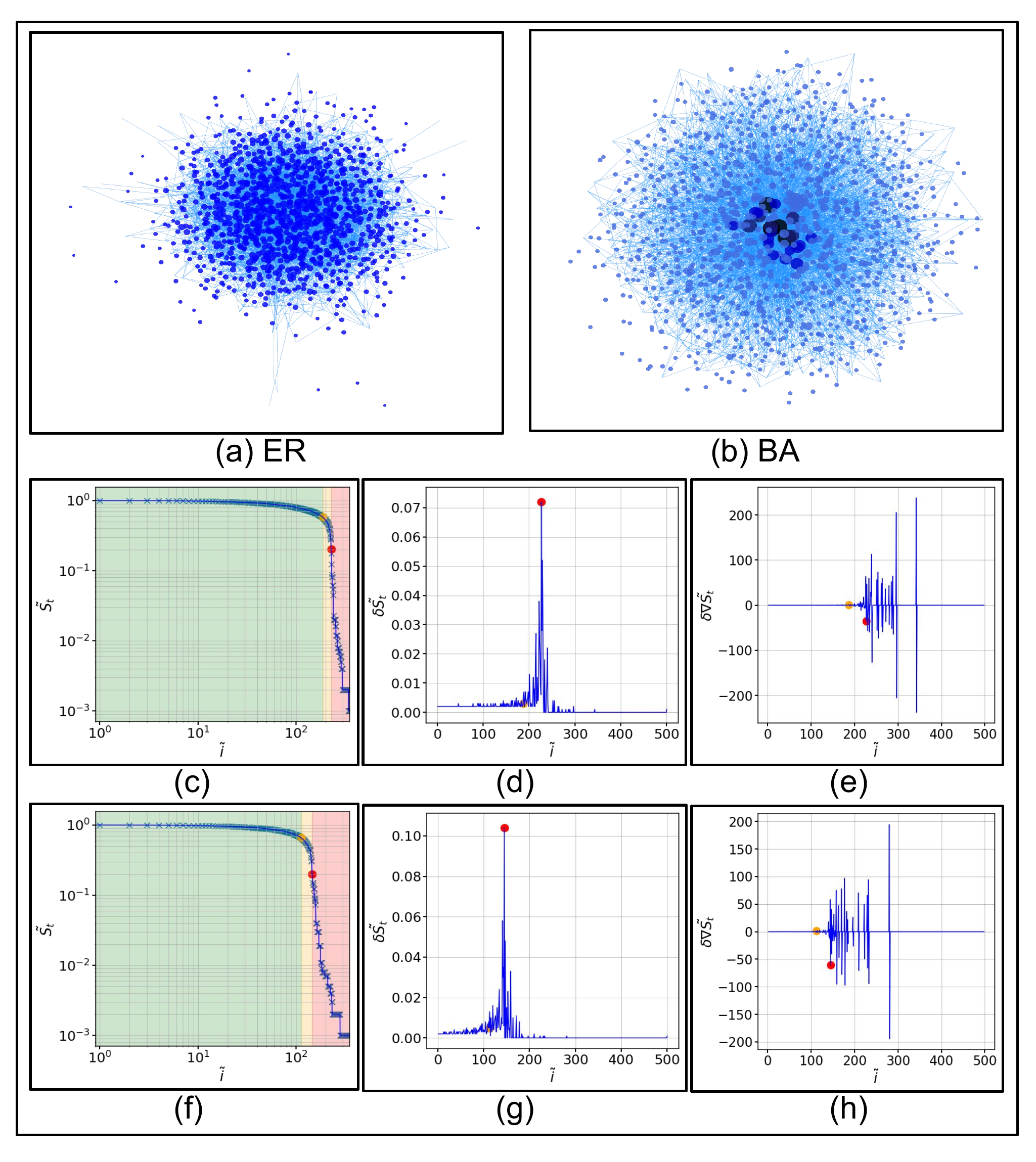}
\caption{\textbf{Delineating functionality regions in synthetic networks}. Same as Fig. 2 but for synthetic networks. For both ER and BA networks, tipping points and onset of warning regions are distinctly visible from  $\Tilde{S_t}$ and $\delta\nabla{\Tilde{S_t}}$ curves.}
\label{fig:3}

\end{figure*}

Further, we identify the warning and tipping points for synthetic networks, ER and BA (Fig. \ref{fig:3}(a) and \ref{fig:3}(b)), under TAM-I and TAM-II discussed above. These two networks consist of an equal number of nodes and edges but have significantly different degree distributions (See SI: Fig. 1). For both systems, we observe warning and tipping points that can be easily delineated using graphical and numerical approaches discussed in the present research. We note that the warning and tipping points for ER appear later in comparison to BA networks for the same disruption size ($B =2$) under the TAM-I disruption scenario. It can be explained by the fact that ER networks lack distinct hubs (nodes with disproportionately large degrees) with significant nodes having comparable degrees, which conforms to the earlier findings reported elsewhere \cite{bollobas2004robustness,morone2015influence} for the gradual removal of nodes.  

\begin{figure*}[htb]
     \centering
     \begin{subfigure}[b]{0.8\textwidth}
         \centering
         \includegraphics[width=\textwidth]{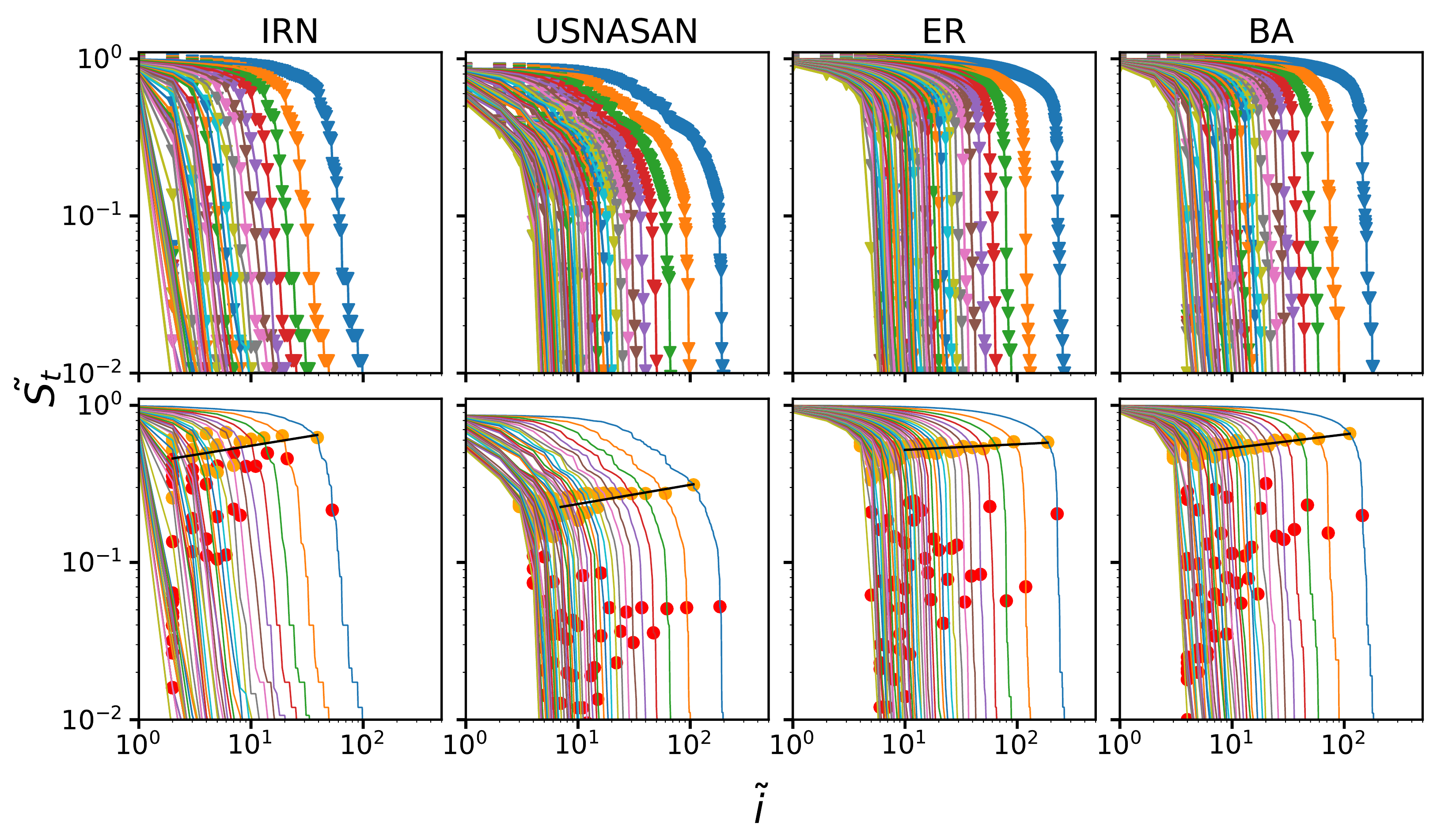}
         \caption{Warning points and tipping points for TAM-I}
         \label{fig:a}
     \end{subfigure}
     \begin{subfigure}[b]{0.8\textwidth}
         \centering
         \includegraphics[width=\textwidth]{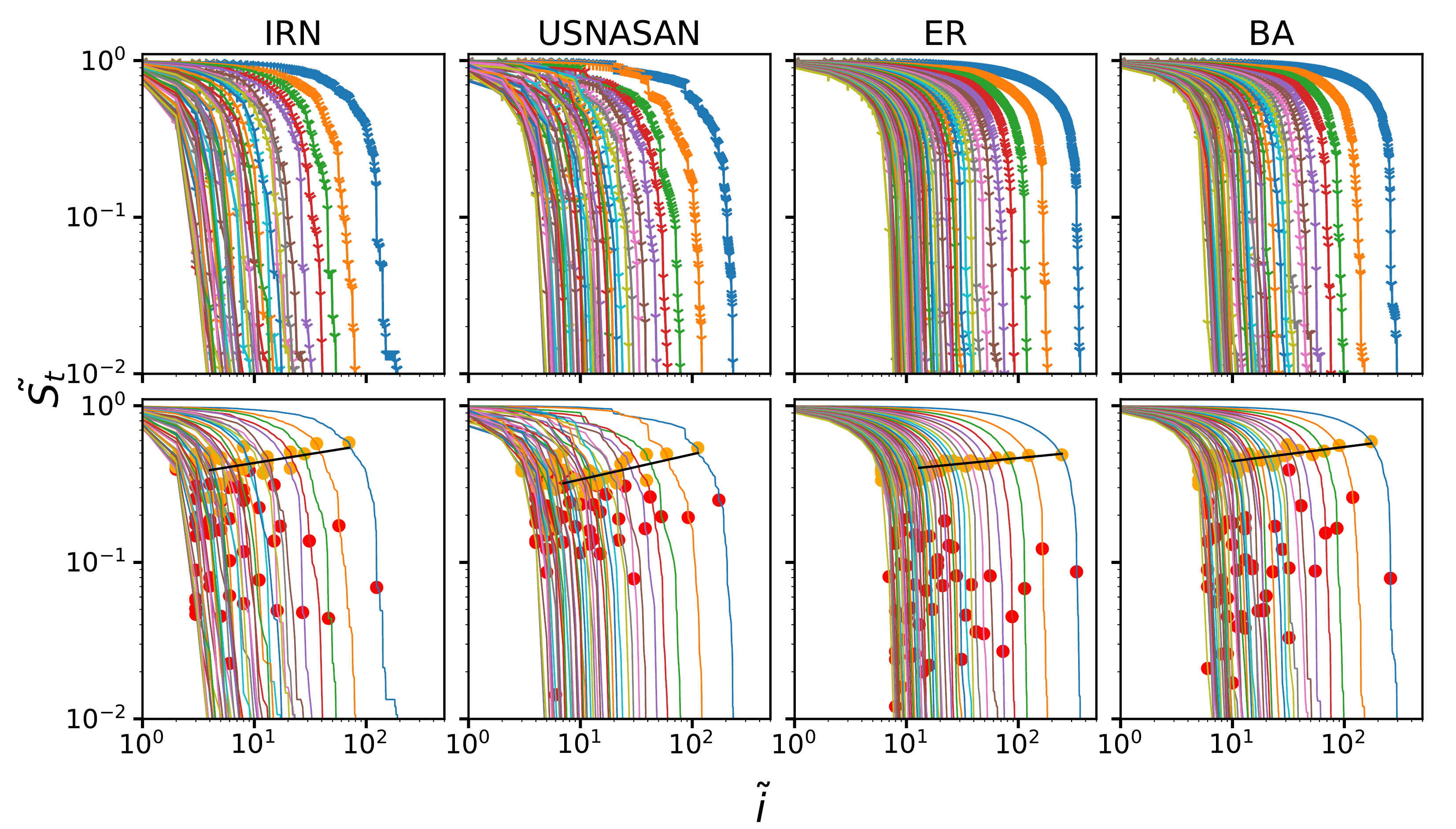}
         \caption{Warning points and tipping points for TAM-II}
         \label{fig:b}
     \end{subfigure}
        \caption{\textbf{Robustness characteristics under disruptions of varying magnitudes } (a) and (b) shows the state of critical functionality ($\Tilde{S_t}$) at every instance for four different networks, namely: IRN, USNASAN, ER (1000 nodes), and BA (1000 nodes). The second row contains the warning point (in orange) and tipping points (in red). \textbf{(a)} corresponds to the targeted attack model (TAM-I) in which the degree of every node is recomputed at every instance. \textbf{((b)} corresponds to an attack based on the probability of the selection of the nodes (TAM-II). Each subplot contains 49 curves corresponding to the 49 batch sizes. In all of these subplots, as we move from left to right, the batch size decreases from 98 to 2 at an interval of 2. It can be observed that the warning points in USNASAN are obtained at lower values of $\Tilde{S_t}$ as compared to the other three because of the high graph density. However, the warning point occurs latest in the case of the ER network because of its randomness and absence of high degree nodes. The warning point for different networks and different attack models follows a particular pattern shown using a black line, which is the best fit line. On the other hand, the tipping points show significant variation for all networks under different perturbation scenarios.}
        \label{fig:4}
\end{figure*}

Fig.  \ref{fig:4}(a) and \ref{fig:4}(b) show the warning and tipping points obtained for values of $B$ ranging from 2 to 98 with an increment of 2 for TAM-I and TAM-II, respectively (see Fig. S3 and Fig. S4 for RAM and TAM-III). Top row of both Fig. \ref{fig:4}(a) and \ref{fig:4}(b) shows the $49$ robustness curves ($\Tilde{S_t}$ vs $\Tilde{i}$), whereas bottom row  shows the warning points and tipping points identified using our approach for two disruption scenarios (See Materials and Methods). The curves on the extreme right in all the panels indicate $B=2$. The value of $B$ increases towards the left with the innermost curve, indicating $B$ of size 98. We note that warning points exhibit a consistent linear pattern on a log-log scale for all disruption scenarios. On the other hand, in the case of tipping points, we do not observe a similar pattern. From  an infrastructure stakeholder perspective,  the ability to predict warning points can significantly help in timely decision making hence averting disproportionate losses. We further notice that the detection of warning points and tipping points becomes challenging
for large values of $B$, making the networks highly vulnerable to disruptions of larger magnitude irrespective of their network properties. 

\begin{figure*}[htb]
\centering
\includegraphics[scale = 0.45]{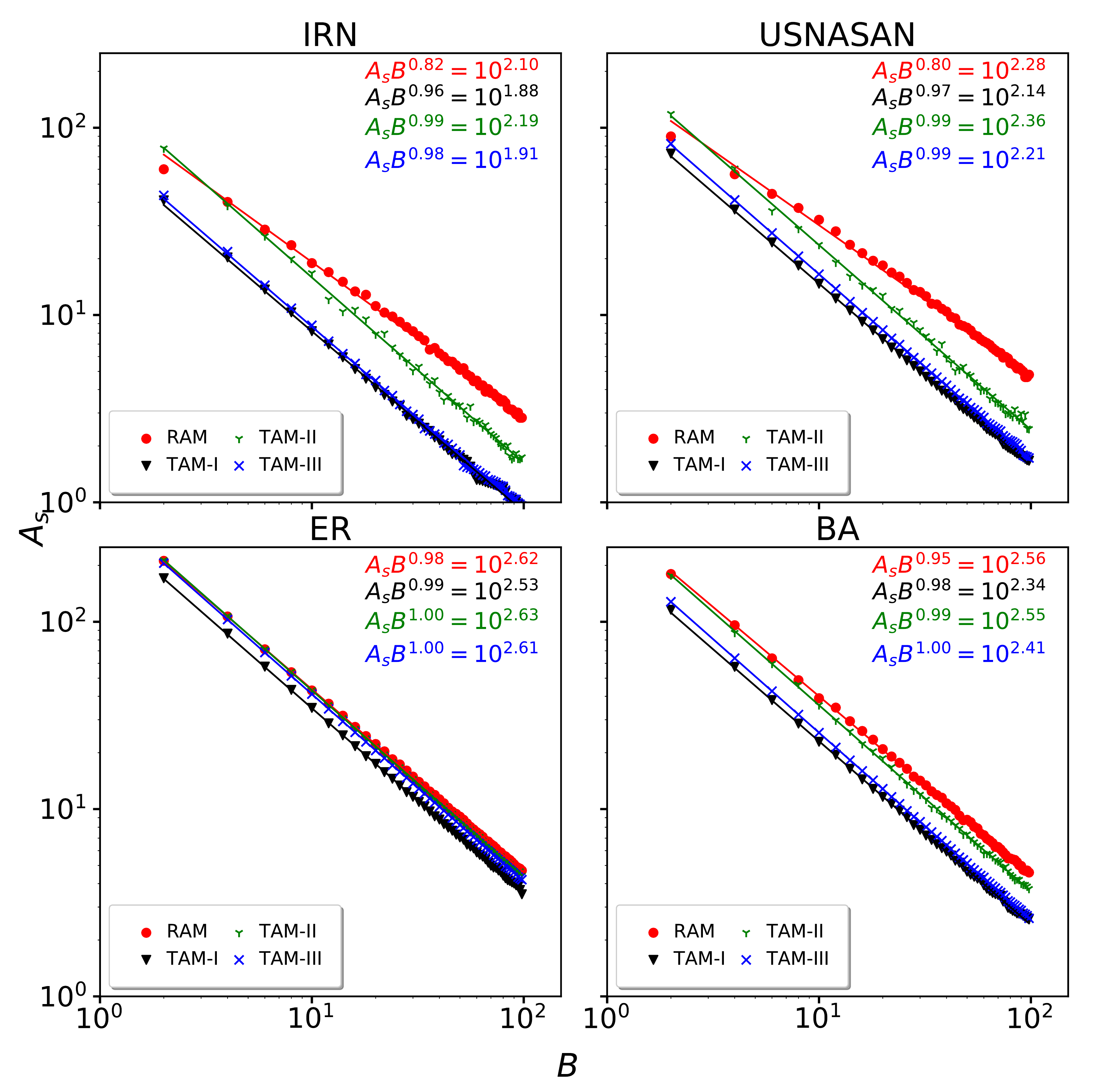}
\caption{\textbf{Robustness characteristics and perturbation size}. Log-log plot of \textit{area} ($A_s$) vs \textit{batch size} ($B$) for the four different networks IRN, USNASAN, ER (1000 nodes), and BA (1000 nodes) with $B$ ranging from 98 to 2 at an interval of 2. Here, $A_s$ corresponds to the area under the contingency curve (graph of $\Tilde{S_t}$ vs. $\Tilde{i}$). The robustness of a network is directly proportional to the $A_s$. Lower the value of $A_s$ for an attack model, higher is the damage. The plot resembles a  straight line. The power of ‘B’ in the equation $A_sB^{m} = C$ represents the slope of the straight line, and the $C$ (a constant) represents the y-intercept of the best fit line. Note that the sequence of lines from top to bottom is always the same, i.e., RAM, TAM-III, TAM-II, and TAM-I. Only the distance between these lines varies.}
\label{fig:5}
\end{figure*}

Finally, we measure the disruption robustness of a network under various attack scenarios and different values of $B$ by measuring the area under the $\Tilde{S_t}$ curve. A larger area indicates more robust network \cite{bhatia2015network,clark2018resilience,tipton2018fungi,linkov2014changing}. We plot the area under the robustness curves ($A_s$) against $B$ on log-log scale for various disruption scenarios (Fig. \ref{fig:5}). For all cases, the relationship between $A_s$ and $B$ is summarized as:
\begin{equation}
 A_s\times B^m = C 
\end{equation}
m is the slope, and C is the y-intercept of the best fit line on the log-log scale (Table S2). For synthetic networks, we also check the asymptotic behavior of the relationship, thus obtained between $A_s$ and $B$ (See Fig.S5). While the value of m is different for various disruption scenarios for IRN,  USNASAN, and BA networks, $m$ approaches unity for ER networks as well as BA networks with a considerably large number of nodes. The $A_s$ and $B$ relationship thus obtained can help us to estimate the robustness characteristics of various networks under the varying magnitude of $B$ merely by visual inspection.

\section*{Discussion}
The first step toward designing and operating networked systems that are robust against attacks and failures is to identify the set of nodes, which, if perturbed, can inflict maximum damage to the network functionality \cite{morone2015influence,barabasi2016network}. In the context of random and idealized systems, the approaches to identify an optimal set of the most critical set of nodes is typically based on topological properties and discussed elsewhere \cite{morone2015influence,barabasi2016network}. However, real-world examples where networked systems are subject to recurrent perturbations of large magnitudes are often reported but seldom studied \cite{buldyrev2010catastrophic,bhatia2015network}. Secondly, from stakeholder and infrastructure managers’ perspective, early warning indicators of the points of sudden collapse or tipping points can help inform graceful degradation \cite{sansavini2017engineering} and hence prevent irreversible losses under sequential disruption scenarios. Although there is growing interest  from a policymaking perspective to proactively ensure critical systems are resilient and maintain mission-critical functions under compromise due to unprecedented events, few analytical tools exist to guide the planning needed to achieve risk minimization and resilience goals.  Generating loss of functionality estimates in infrastructure networks and cascading failure impacts will inform consequence-based decision support, risk assessments, and contingency analysis using mitigation strategies. Moreover, increasing the speed and precision in estimating loss of functionality through the warning region and tipping point analysis here will fundamentally improve pre- and post-event decision making for infrastructure stakeholders thereby enhancing national security and economic prosperity.

In this research, using a suite of graphical and numerical methods, we propose an approach to identify such early indicators for networked systems, both real and synthetic, under sequential disruptions. Our results show that despite the inherent differences in the network attributes, such as network density and degree distribution, we can identify consistent patterns in warning regions before disproportionate losses set in when multiple network components are removed simultaneously. 
Our results also illustrate the predictive and surprisingly  straightforward analytical relationship between robustness measure and disruption size for different networks. Hence, combining the early-warning indicators, tipping points, and robustness measures obtained as a function of disruption size can yield predictive insights into the response of networked systems to disruptions of substantial sizes. 

Future extensions to the proposed framework may need to consider the underlying dynamical mechanisms \cite{jiang2018predicting}, and explore how the inclusion of interaction strengths in both dynamical and static models can alter the regions of functionality. Further, we have only considered the node removal, which is too drastic a disruption. Also, in our synthetic disruption scenarios, we assume size of $B$ to be fixed throughout the disruption. However, in real-world, these disruptions can exhibit large variations. From stakeholders’ perspectives, analyzing the simultaneous rewiring or deletion of multiple edges can help further understand effects on  system robustness \cite{kovacs2015network}.

\bibliographystyle{unsrt}  
\bibliography{references}  

\begin{thebibliography}{10}

\bibitem{pocock2012robustness}
Michael~JO Pocock, Darren~M Evans, and Jane Memmott.
\newblock The robustness and restoration of a network of ecological networks.
\newblock {\em Science}, 335(6071):973--977, 2012.

\bibitem{luke2010power}
Timothy~W Luke.
\newblock Power loss or blackout: The electricity network collapse of august
  2003 in north america.
\newblock {\em Disrupted Cities: When Infrastructure Fails}, pages 55--68,
  2010.

\bibitem{ganin2017resilience}
Alexander~A Ganin, Maksim Kitsak, Dayton Marchese, Jeffrey~M Keisler, Thomas
  Seager, and Igor Linkov.
\newblock Resilience and efficiency in transportation networks.
\newblock {\em Science advances}, 3(12):e1701079, 2017.

\bibitem{yazdani2012applying}
Alireza Yazdani and Paul Jeffrey.
\newblock Applying network theory to quantify the redundancy and structural
  robustness of water distribution systems.
\newblock {\em Journal of Water Resources Planning and Management},
  138(2):153--161, 2012.

\bibitem{buldyrev2010catastrophic}
Sergey~V Buldyrev, Roni Parshani, Gerald Paul, H~Eugene Stanley, and Shlomo
  Havlin.
\newblock Catastrophic cascade of failures in interdependent networks.
\newblock {\em Nature}, 464(7291):1025--1028, 2010.

\bibitem{wang2014robustness}
Jianwei Wang, Chen Jiang, and Jianfei Qian.
\newblock Robustness of internet under targeted attack: a cascading failure
  perspective.
\newblock {\em Journal of Network and Computer Applications}, 40:97--104, 2014.

\bibitem{jackson2001historical}
Jeremy~BC Jackson, Michael~X Kirby, Wolfgang~H Berger, Karen~A Bjorndal,
  Louis~W Botsford, Bruce~J Bourque, Roger~H Bradbury, Richard Cooke, Jon
  Erlandson, James~A Estes, et~al.
\newblock Historical overfishing and the recent collapse of coastal ecosystems.
\newblock {\em science}, 293(5530):629--637, 2001.

\bibitem{svendsen2007connectivity}
Nils~Kalstad Svendsen and Stephen~D Wolthusen.
\newblock Connectivity models of interdependency in mixed-type critical
  infrastructure networks.
\newblock {\em Information Security Technical Report}, 12(1):44--55, 2007.

\bibitem{yletyinen2019understanding}
Johanna Yletyinen, Philip Brown, Roger Pech, Dave Hodges, Philip~E Hulme,
  Thomas~F Malcolm, Fleur~JF Maseyk, Duane~A Peltzer, George~LW Perry, Sarah~J
  Richardson, et~al.
\newblock Understanding and managing social--ecological tipping points in
  primary industries.
\newblock {\em BioScience}, 69(5):335--347, 2019.

\bibitem{dakos2019ecosystem}
Vasilis Dakos, Blake Matthews, Andrew~P Hendry, Jonathan Levine, Nicolas
  Loeuille, Jon Norberg, Patrik Nosil, Marten Scheffer, and Luc De~Meester.
\newblock Ecosystem tipping points in an evolving world.
\newblock {\em Nature ecology \& evolution}, 3(3):355--362, 2019.

\bibitem{lever2014sudden}
J~Jelle Lever, Egbert~H van Nes, Marten Scheffer, and Jordi Bascompte.
\newblock The sudden collapse of pollinator communities.
\newblock {\em Ecology letters}, 17(3):350--359, 2014.

\bibitem{gao2011robustness}
Jianxi Gao, Sergey~V Buldyrev, Shlomo Havlin, and H~Eugene Stanley.
\newblock Robustness of a network of networks.
\newblock {\em Physical Review Letters}, 107(19):195701, 2011.

\bibitem{dong2013robustness}
Gaogao Dong, Jianxi Gao, Ruijin Du, Lixin Tian, H~Eugene Stanley, and Shlomo
  Havlin.
\newblock Robustness of network of networks under targeted attack.
\newblock {\em Physical Review E}, 87(5):052804, 2013.

\bibitem{jiang2018predicting}
Junjie Jiang, Zi-Gang Huang, Thomas~P Seager, Wei Lin, Celso Grebogi, Alan
  Hastings, and Ying-Cheng Lai.
\newblock Predicting tipping points in mutualistic networks through dimension
  reduction.
\newblock {\em Proceedings of the National Academy of Sciences},
  115(4):E639--E647, 2018.

\bibitem{gao2016universal}
Jianxi Gao, Baruch Barzel, and Albert-L{\'a}szl{\'o} Barab{\'a}si.
\newblock Universal resilience patterns in complex networks.
\newblock {\em Nature}, 530(7590):307--312, 2016.

\bibitem{morone2019k}
Flaviano Morone, Gino Del~Ferraro, and Hern{\'a}n~A Makse.
\newblock The k-core as a predictor of structural collapse in mutualistic
  ecosystems.
\newblock {\em Nature physics}, 15(1):95--102, 2019.

\bibitem{jiang2019harnessing}
Junjie Jiang, Alan Hastings, and Ying-Cheng Lai.
\newblock Harnessing tipping points in complex ecological networks.
\newblock {\em Journal of the Royal Society Interface}, 16(158):20190345, 2019.

\bibitem{albert2000error}
R{\'e}ka Albert, Hawoong Jeong, and Albert-L{\'a}szl{\'o} Barab{\'a}si.
\newblock Error and attack tolerance of complex networks.
\newblock {\em nature}, 406(6794):378--382, 2000.

\bibitem{barabasi2016network}
Albert-L{\'a}szl{\'o} Barab{\'a}si et~al.
\newblock {\em Network science}.
\newblock Cambridge university press, 2016.

\bibitem{bhatia2015network}
Udit Bhatia, Devashish Kumar, Evan Kodra, and Auroop~R Ganguly.
\newblock Network science based quantification of resilience demonstrated on
  the indian railways network.
\newblock {\em PloS one}, 10(11):e0141890, 2015.

\bibitem{bollobas2004robustness}
B{\'e}la Bollob{\'a}s and Oliver Riordan.
\newblock Robustness and vulnerability of scale-free random graphs.
\newblock {\em Internet Mathematics}, 1(1):1--35, 2004.

\bibitem{steinberg2011baton}
Laura Steinberg, Nicholas Santella, and Corri Zoli.
\newblock Baton rouge post-katrina: the role of critical infrastructure
  modeling in promoting resilience.
\newblock {\em Homeland Security Affairs}, 7, 2011.

\bibitem{gariel2008graceful}
Maxime Gariel and Eric Feron.
\newblock Graceful degradation of air traffic operations: airspace sensitivity
  to degraded surveillance systems.
\newblock {\em Proceedings of the IEEE}, 96(12):2028--2039, 2008.

\bibitem{scheffer2012anticipating}
Marten Scheffer, Stephen~R Carpenter, Timothy~M Lenton, Jordi Bascompte,
  William Brock, Vasilis Dakos, Johan Van~de Koppel, Ingrid~A Van~de Leemput,
  Simon~A Levin, Egbert~H Van~Nes, et~al.
\newblock Anticipating critical transitions.
\newblock {\em science}, 338(6105):344--348, 2012.

\bibitem{clark2018resilience}
Kevin~L Clark, Udit Bhatia, Evan~A Kodra, and Auroop~R Ganguly.
\newblock Resilience of the us national airspace system airport network.
\newblock {\em IEEE Transactions on Intelligent Transportation Systems},
  19(12):3785--3794, 2018.

\bibitem{molloy1995critical}
Michael Molloy and Bruce Reed.
\newblock A critical point for random graphs with a given degree sequence.
\newblock {\em Random structures \& algorithms}, 6(2-3):161--180, 1995.

\bibitem{barabasi2003scale}
Albert-L{\'a}szl{\'o} Barab{\'a}si and Eric Bonabeau.
\newblock Scale-free networks.
\newblock {\em Scientific american}, 288(5):60--69, 2003.

\bibitem{albert2000topology}
R{\'e}ka Albert and Albert-L{\'a}szl{\'o} Barab{\'a}si.
\newblock Topology of evolving networks: local events and universality.
\newblock {\em Physical review letters}, 85(24):5234, 2000.

\bibitem{pagani2013power}
Giuliano~Andrea Pagani and Marco Aiello.
\newblock The power grid as a complex network: a survey.
\newblock {\em Physica A: Statistical Mechanics and its Applications},
  392(11):2688--2700, 2013.

\bibitem{reka2000internet}
A~Reka et~al.
\newblock The internet achilles' heel: Error and attack tolerance of complex
  networks.
\newblock {\em Physica A, Elsevier Science bv}, 2000.

\bibitem{borgatti2006robustness}
Stephen~P Borgatti, Kathleen~M Carley, and David Krackhardt.
\newblock On the robustness of centrality measures under conditions of
  imperfect data.
\newblock {\em Social networks}, 28(2):124--136, 2006.

\bibitem{xie2019eradicating}
Jiarong Xie, Youyou Yuan, Zhengping Fan, Jiahai Wang, Jiajing Wu, and Yanqing
  Hu.
\newblock Eradicating abrupt collapse on single network with dependency groups.
\newblock {\em Chaos: An Interdisciplinary Journal of Nonlinear Science},
  29(8):083111, 2019.

\bibitem{morone2015influence}
Flaviano Morone and Hern{\'a}n~A Makse.
\newblock Influence maximization in complex networks through optimal
  percolation.
\newblock {\em Nature}, 524(7563):65--68, 2015.

\bibitem{tipton2018fungi}
Laura Tipton, Christian~L M{\"u}ller, Zachary~D Kurtz, Laurence Huang, Eric
  Kleerup, Alison Morris, Richard Bonneau, and Elodie Ghedin.
\newblock Fungi stabilize connectivity in the lung and skin microbial
  ecosystems.
\newblock {\em Microbiome}, 6(1):12, 2018.

\bibitem{linkov2014changing}
Igor Linkov, Todd Bridges, Felix Creutzig, Jennifer Decker, Cate Fox-Lent,
  Wolfgang Kr{\"o}ger, James~H Lambert, Anders Levermann, Benoit Montreuil,
  Jatin Nathwani, et~al.
\newblock Changing the resilience paradigm.
\newblock {\em Nature Climate Change}, 4(6):407, 2014.

\bibitem{sansavini2017engineering}
Giovanni Sansavini.
\newblock Engineering resilience in critical infrastructures.
\newblock In {\em Resilience and Risk}, pages 189--203. Springer, 2017.

\bibitem{kovacs2015network}
Istv{\'a}n~A Kov{\'a}cs and Albert-L{\'a}szl{\'o} Barab{\'a}si.
\newblock Network science: Destruction perfected.
\newblock {\em Nature}, 524(7563):38--39, 2015.

\end{thebibliography}

\section*{Acknowledgments}
This work is supported by Internal Research Grant, Indian Institute of Technology, Gandhinagar, and Startup Research Grant, Science and Engineering Research Board.  We are thankful to Prof. Auroop Ratan Ganguly, Northeastern University, Boston and Divya Upadhyay, Indian Institute of Technology, Gandhinagar for their helpful comments and suggestions.

\textbf{Author contributions:} UB, PP, and MS designed the experiments, UG, and DK performed the analysis. UB, UG, MS, and SC wrote the manuscript. PP performed the mathematical analysis.

\textbf{Competing interests:}Authors declare no conflict of interest.

\section*{Materials and Methods}
\subsection*{Datasets}
This paper uses four undirected network datasets - two synthetic networks and two real-world networks. The two synthetic networks comprise classical Erdős–Rényi (ER) and Barab\'{a}si–Albert (BA) networks of different sizes. Whereas the Indian Railways Network (IRN) and the U.S. National Airspace System Airport Network (USNASAN) comprise two real-world networks. In IRN, we study origin-destination data of passenger-carrying trains. This network was constructed using the data cconstructed in \cite{bhatia2015network} comprising 809 nodes \textbf{or stations} and 2,342 edges \textbf{or links between these stations}. We have modelled the IRN as an origin-destination network where we consider stations with at least one originating or terminating train. This large network of 809 stations comprises of 7,066 trains as of October 30, 2014. Only 752 of the 809 stations are part of the giant component or the largest connected group of stations.Similarly, we model USNASAN after \cite{clark2018resilience}. USNASAN comprises of 1261 nodes with 47489 edges with all nodes as part of the largest connected cluster at $t=0$.

For experimental consistency, we design synthetic networks (ER and BA) with equal number of nodes and fix the graph density approximately equal to the graph density of the largest cluster of IRN. Here, the graph density ($D$) for an undirected graph with $|V|$ nodes and $|E|$ edges is defined as
\begin{equation}
    D = \frac{2|E|}{|V|(|V|-1)}
\end{equation}

The ER and BA networks were constructed using the Python's NetworkX library\footnote{\url{https://networkx.github.io/}}. ER network with 1,000 nodes and graph density 0.008 was constructed. ER networks with different number of nodes (upto 25,000 nodes) and same graph density were also constructed for more detailed analysis. Similarly, a BA network of 1,000 nodes along with networks containing different number of nodes (upto 25,000 nodes) but almost same graph density were constructed for analysis. 

\begin{table}[!tbh]
\centering
\caption{Data Statistics}
\begin{tabular}{lrrrrr}
Network Name & Nodes & Edges & No. of components & Size Of Giant Component  \\
\midrule
ER &  1000 & 3981 & 1 & 1000 \\
BA &  1000   &  3984     &  1 & 1000 \\
IRN &  809  &   2342   &    27 & 752   \\
USNASAN &  1261   &  47489   &    1 & 1261 \\
\bottomrule
\end{tabular}
\end{table}

\subsection*{Determining Network Robustness}
Infrastructure network failure and disruptions are inevitable, particularly due to both internal and external causes such as security breaches, natural calamities, node aging, etc. Different networks respond to similar type of failures differently due to underlying structural properties. In a network, robustness refers to the system's ability to carry out its basic functions even when some nodes/links have been failed~\cite{barabasi2016network}. Thus, network structure results in differential robustness. The failures are classified into two broad categories: (i) random, and (ii) targeted. Random failures occur when nodes fail without any ordering, i.e., disruptions  cannot be predicted with certainty. Targeted failures occurs when nodes fail in a prescribed ordering, i.e., disruptions  can be predicted with some certainty. In this work, we define a continuous 
failure regime, where few nodes in the networks fail at each time-step without recovery. The \textit{network critical functionality} ($S$) is defined in terms of  giant component that exist in the original network (at time-step  $t=0$).  The network critical functionality at each time-step $t$ ($\Tilde{S_t}$) as the ratio of remaining number of nodes in the giant component at time-step $t$ (fragmented functionality) to the number of stations in the initial giant component (total functionality)~\cite{bhatia2015network}.
\begin{equation}
         \Tilde{S_t} = \frac{Fragmented\ Functionality}{Total\ Functionality\ }
\end{equation}

\subsection*{The contingency curve}
The contingency curve presents an visualization of  network critical functionality ($\Tilde{S_t}$) at each time-step. Specifically, we remove $B$ nodes at each time-step or instance ($\Tilde{i}$) and compute $\Tilde{S_t}$.  In the current experimentation, generate $K$  =100 sets of size $B$  for the random attack model. The higher values of $K$, even though are computationally expensive, did not yield significant improvements.  Figure~\ref{fig:warning} presents an illustrative example of contingency curve on log-log scale. Note that, we consider log-log scale for better visualization of lower $\Tilde{S_t}$ values.  The curve also presents an insight of warning points and tipping points. We define \textit{`tipping point'} as that time-step at which the network drastically looses its functionality.  A delayed tipping point results in a more robust network.

\paragraph{\textbf{Identifying warning point}}
Figure~\ref{fig:warning} also presents the methodology to identify warning point in the contingency curve. The initial step to join two distinct points on the curve with a line (say $l_1$). The first point lies at $\Tilde{i}$ = 1 and the second point lies just before the point with $\Tilde{S}$ = 0. In the next step, we draw parallel lines ($l_2,l_3,...,l_k$) towards right of the $l_1$. We stop at line $l_k$ such that it touches only one of the point on the curve and has the highest value of y-intercept. We term this point as \textit{`warning point'}. 

\paragraph{\textbf{Identifying tipping point}}
We use the contingency curve and plot the deviation in the size of largest connected component ($\delta{\Tilde{S_t}}$). The point with maximum deviation in the size of largest connected component, after the warning point, is the required tipping point.

\paragraph{\textbf{Area under contingency curve}}
Area under contingency curve indicates the robustness of the network under continuous attack regime. We study the relation between area under contingency curve $A_S$ against batch size $B$. Simple curve fitting experiments suggest a rectangular hyperbolic nature of the curve. Note that, we fit the log-log scaled $A_S$ and $B$ with a straight line $A_SB^{-m}\ =\ C$.

\section*{Theoretical analysis of tipping point} 
In Fig.~\ref{fig:warning}, a method to identify the tipping point for a network under attack (uniform random or preferential) is explained. Here the given pictorial analysis is formulated to analyse the behaviour of tipping point under different criteria is explained.  In Fig.~\ref{fig:warning}, horizontal axis represents time and each time step a node get removed with some probability $p'$, which can be transferred as occupation probability. Let $\phi$ be the occupation probability then $1-\phi=active\ nodes/total\ nodes= tp'/n=x$. As we know that tipping point is the intersection of non-linear curve (in blue cross marks) and a line parallel to the line joining the first and last points of the  non-linear curve (in blue cross marks). Thus to analyse the behaviour of tipping point, we need equation of the line and non-linear curve. Solving them for single point, we get the tipping point.

First point A of the curve is $(\ln{\phi},\ln{\Tilde{S_x}})=(0,0)$ and last point B is $(\ln{\phi_c},\ln{\frac{1}{n}})$, the equation the line passing through points A and B is \begin{equation}
    \ln{\Tilde{S_x}}=\frac{\ln{\frac{1}{n}}}{\ln{\phi_c}}\ln{\phi}+\mathcal{C},
    \label{line}
\end{equation}
where $\phi_c$ is critical occupation probability and $1-x=\phi$. Now, we need the equation of the curve which is corresponding to size of giant component. 

Consider a random network having degree distribution $p_k$, and it is under uniform random attack. Let $\Tilde{S_x}$ is the size of giant component when occupation probability is $\phi=1-x$. Let a node have degree $k$ and it is not connected to the giant component with average probability $u$ via a particular neighbouring node. The average probability that a node is not connected to giant component is $$\overline{u}=\sum_kp_ku^k.$$

Hence, the size of the giant component is defined by 
\begin{equation}
    \Tilde{S_x}=\phi(1-\overline{u}).
    \label{giant}
\end{equation}

\textbf{How to compute $u$:} As we know that $u$ is the average probability that a node is not connected to giant component via a particular neighbouring node let say X. This event can happen in two ways: Either X is not present that has probability $(1-\phi)$, or X is present but X is not connected to giant component via its remaining neighbours. Thus, the combined probability that a node is not connect to giant cluster via X is $1-\phi+\phi u^k$, where $k$ is distributed according to excess degree distribution defined by $$q_k=\frac{(k+1)p_{k+1}}{\overline{k}}.$$ $\overline{k}$ is average degree in the network. Thus, 

\begin{equation}
    u=1-\phi+\phi\sum_kq_ku^k=1-\phi+\phi\overline{v},
    \label{prob0}
\end{equation}
where $\overline{v}=\sum_kq_ku^k.$

Next, $\phi_c=\frac{\overline{k}}{\overline{k^2}-\overline{k}},$ where $\overline{s}$ signifies the average of $s$.
 Assume that $\overline{v}$ is well behaved near $u=1$, its all derivatives exist, then 
 \begin{equation}
     \overline{v}=\overline{v}(1)+(u-1)\overline{v}'(1)+\frac{1}{2}(u-1)^2\overline{v}''(1)+O(u-1)^3=1+\frac{u-1}{\phi_c}+\frac{1}{2}(u-1)^2\overline{v}''(1)+O(u-1)^3.
     \label{prob1}
 \end{equation}

From Eqs.~(\ref{prob0}) and (\ref{prob1}),

\begin{equation}
    u-1=\frac{2}{\overline{v}''(1)}\frac{\phi_c-\phi}{\phi_c\phi}+O(u-1)^2.
    \label{prob2}
\end{equation}

Similarly,

\begin{equation}
    \overline{u}=\overline{u}(1)+(u-1)\overline{u}'(1)+O(u-1)^2=1+\frac{2\overline{k}}{\overline{v}''(1)}\frac{\phi_c-\phi}{\phi_c\phi}+O(\phi-\phi_c)^2.
    \label{prob3}
\end{equation}

From Eqs.~(\ref{giant}) and (\ref{prob3}),

\begin{equation}
    \Tilde{S}_x=\frac{2\overline{k}}{\overline{v}''(1)}\frac{\phi-\phi_c}{\phi_c}+O(\phi-\phi_c)^2.
    \label{giant1}
\end{equation}

\begin{equation}
    \ln{\Tilde{S}_x}=\ln{\frac{2\overline{k}}{\overline{v}''(1)}}+\ln{\frac{\phi-\phi_c}{\phi_c}}  \label{giant2}
\end{equation}






\begin{figure*}[htb]
\centering
\includegraphics[scale = 1.2]{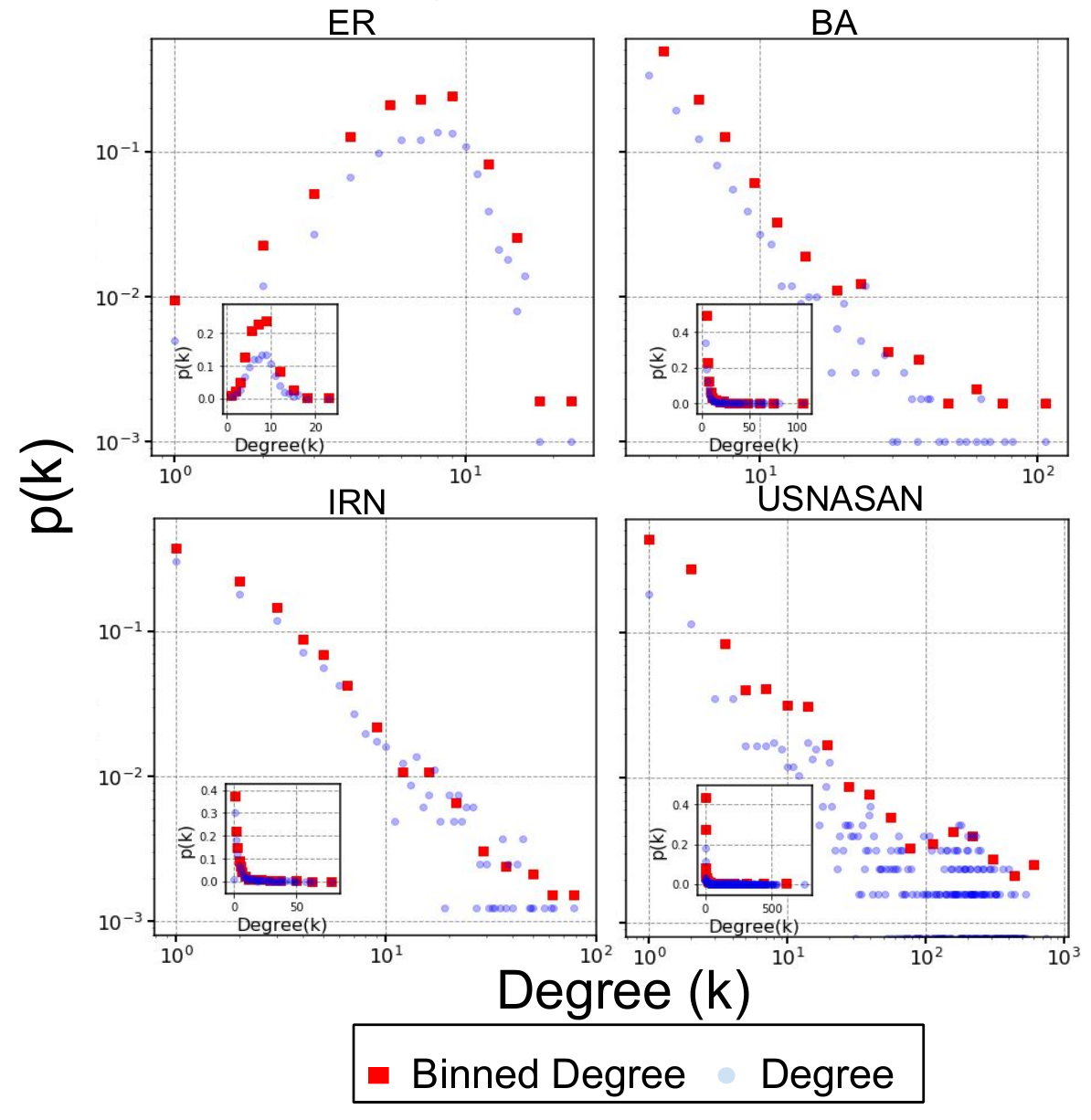}
\caption{\textbf{Degree Distribution:} Degree Distribution of Erdos-Renyi Network (1000 nodes), Barabasi-Albert Network (1000 nodes), Indian Railways Network (809 nodes) and U.S. National Airspace System Airport Network (1261 nodes)  on a Log-log scale. The subplot in each plot is the degree distribution on a normal scale. Degree Distribution of Erdos-Renyi (ER) or Random Network is observed to follow Poisson distribution. Whereas, the other networks follow the power law ($k^{-\alpha}$). For BA network, $\alpha$ = 2.9643 and $\sigma$ = 0.062117 and for IRN, $\alpha$ = 2.075069 and $\sigma$ = 0.045471. Here $\sigma$ is the standard error.}
\end{figure*}

\begin{figure*}[htb]
\centering
\includegraphics[scale = 0.55]{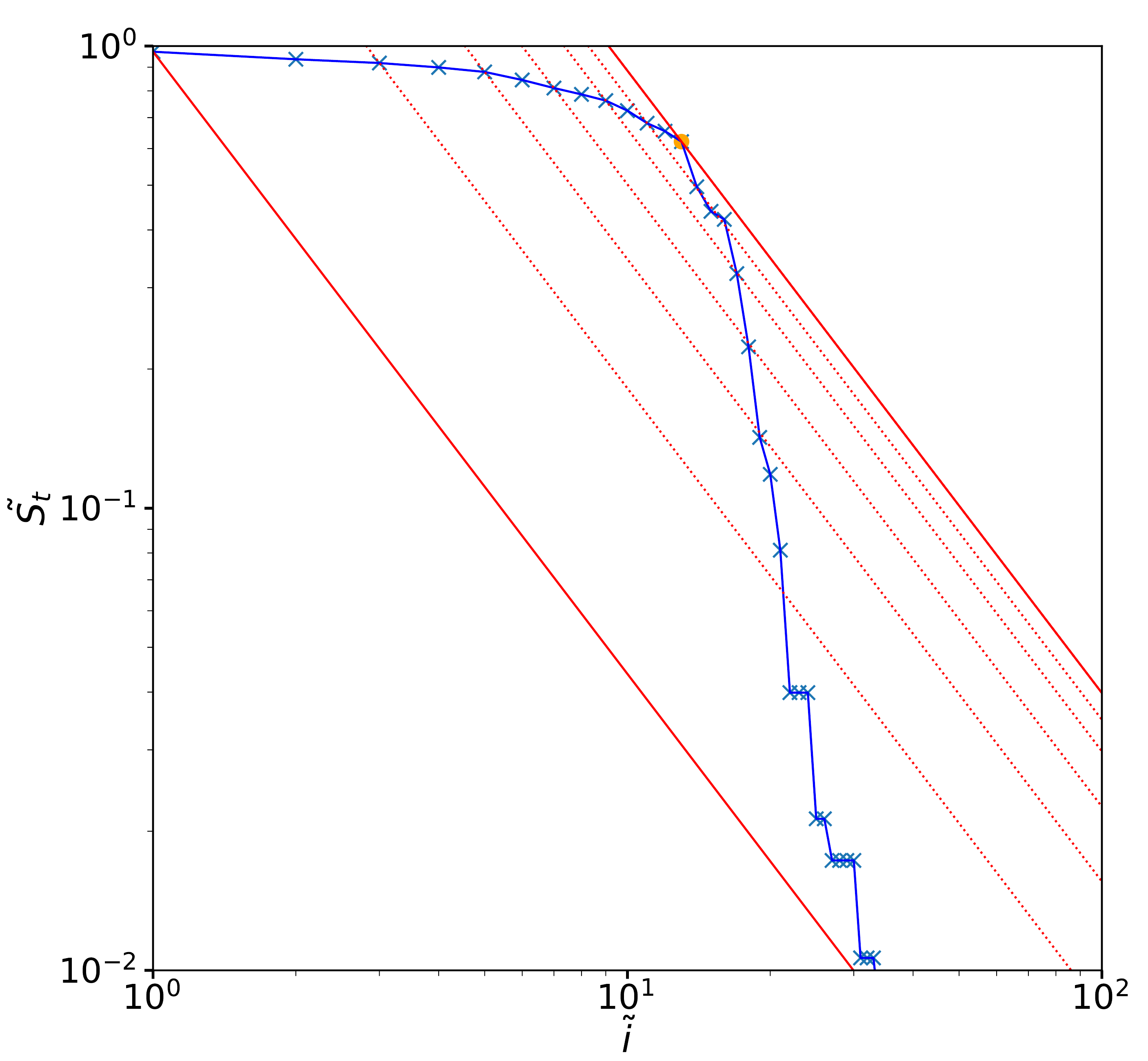}
\caption{\textbf{The Warning Point}. The method used to find the warning point for any network has been depicted in this figure. The log-log plot of the \textit{state of critical functionality ($\Tilde{S_t}$) vs. time/instance  ($\Tilde{i}$)} is used to identify the warning point. Initially, the first point and the last point on $\Tilde{S_t}$ vs. $\Tilde{i}$ curve are joined with a line and the value of slope and y-intercept of the line are obtained. Now, lines parallel to this line are plotted passing through different points. The y-intercept of each line is then obtained and one with the maximum value of y-intercept is then selected. The point through which this line passes is the warning point of the network.}\label{fig:warning}
\end{figure*}

\begin{figure*}[htb]
\centering
\includegraphics[width=\linewidth]{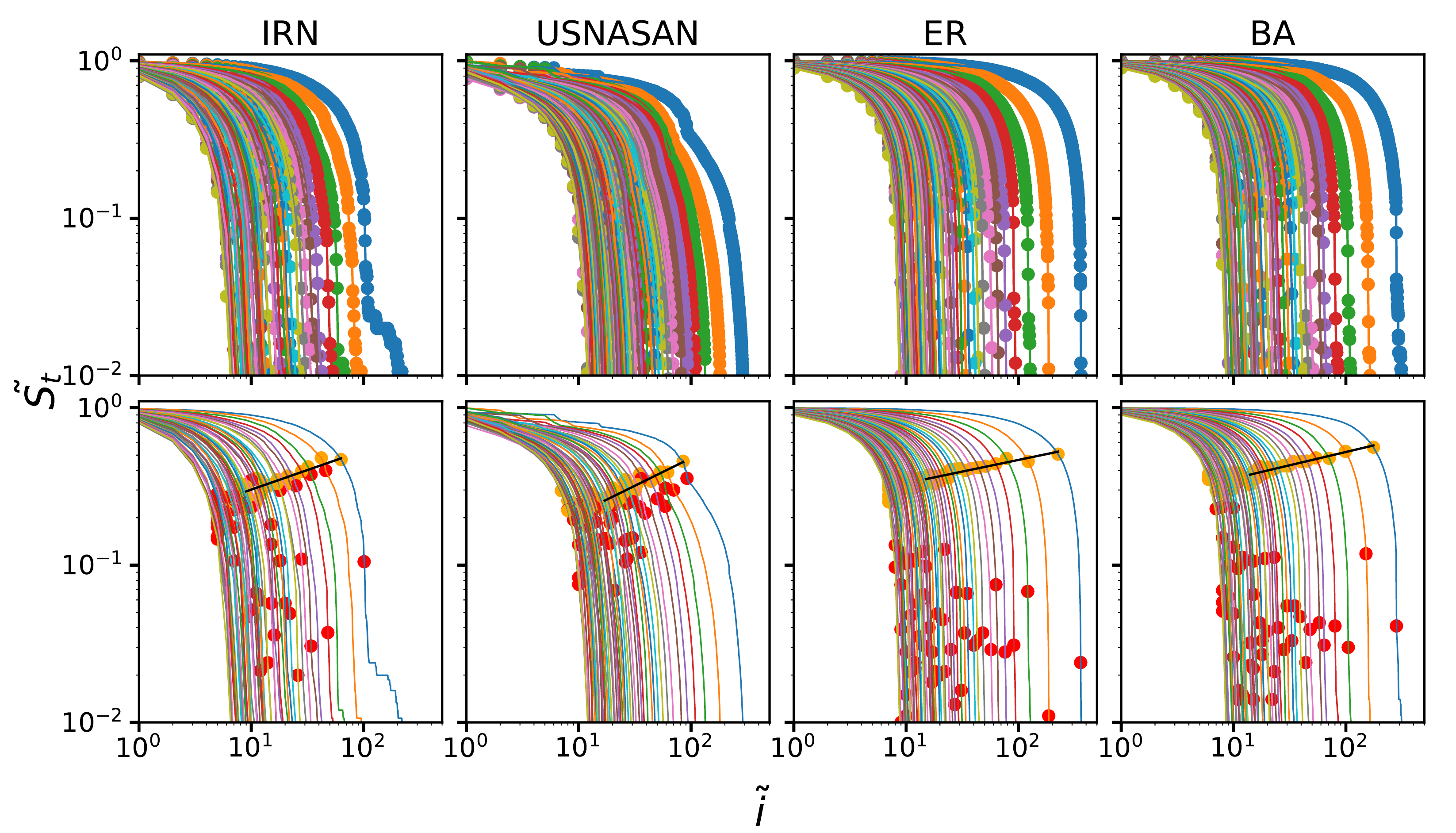}
\caption{\textbf{Warning points and tipping points for RAM}}
\end{figure*}

\begin{figure*}[htb]
\centering
\includegraphics[width=\linewidth]{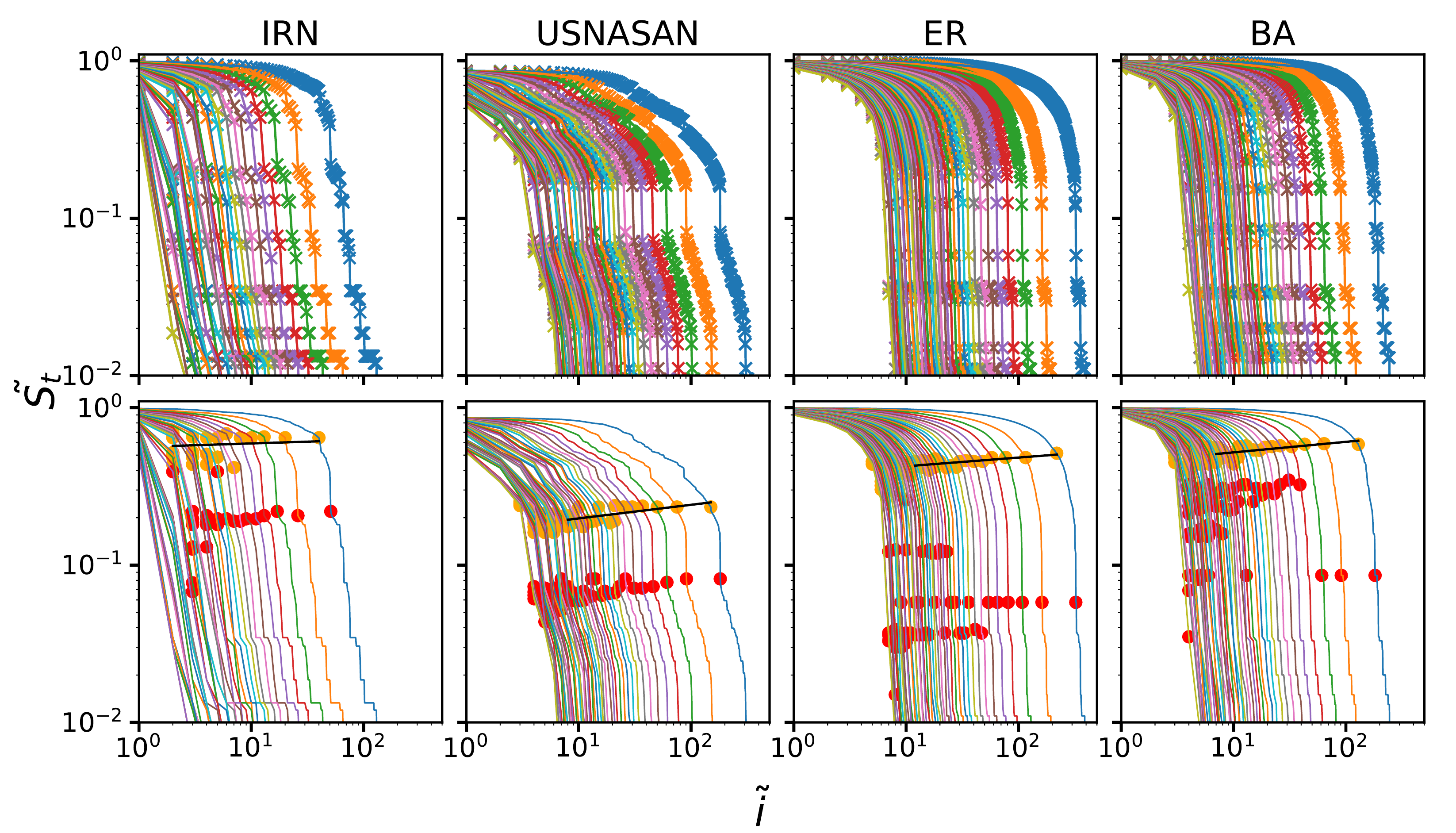}
\caption{\textbf{Warning points and tipping points for TAM-III}}
\end{figure*}

\begin{figure*}[htb]
\centering
\includegraphics[scale = 0.35]{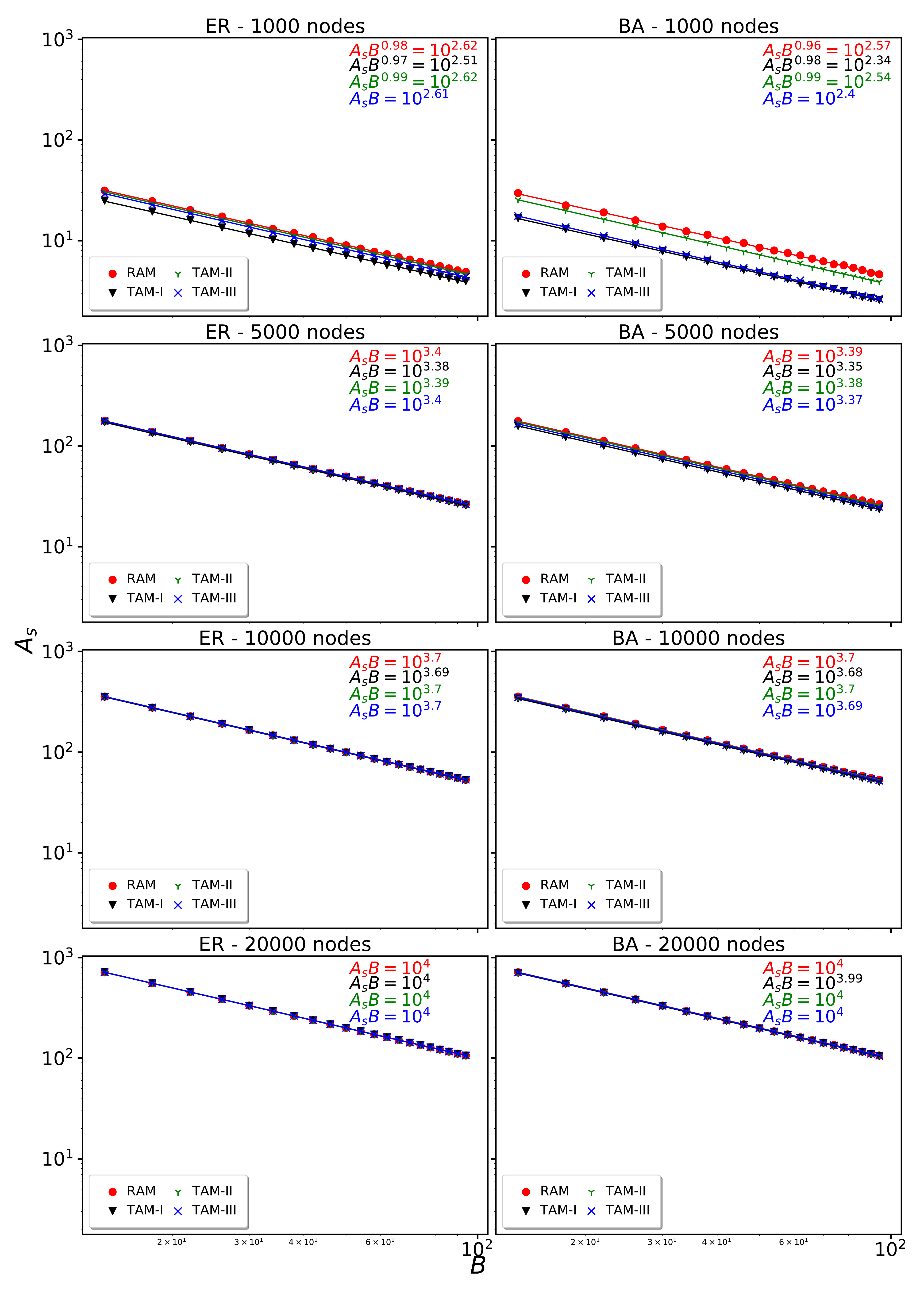}
\caption{\textbf{Large-Scale Networks}. The Area ($A_s$) vs Batch size ($B$) has been plotted in which the batch sizes are 14, 18, 22, $\ldots$, 98. ER and BA network with number of nodes equal to 1000, 5000, 10000 and 20000 were disrupted using all four attack models. It is observed that the distance between these lines slowly decreases. The sequence of the lines from top to bottom is always same. For large networks, the equation of lines also becomes the same.}
\end{figure*}

\begin{figure*}[htb]
\centering
\includegraphics[scale = 0.4]{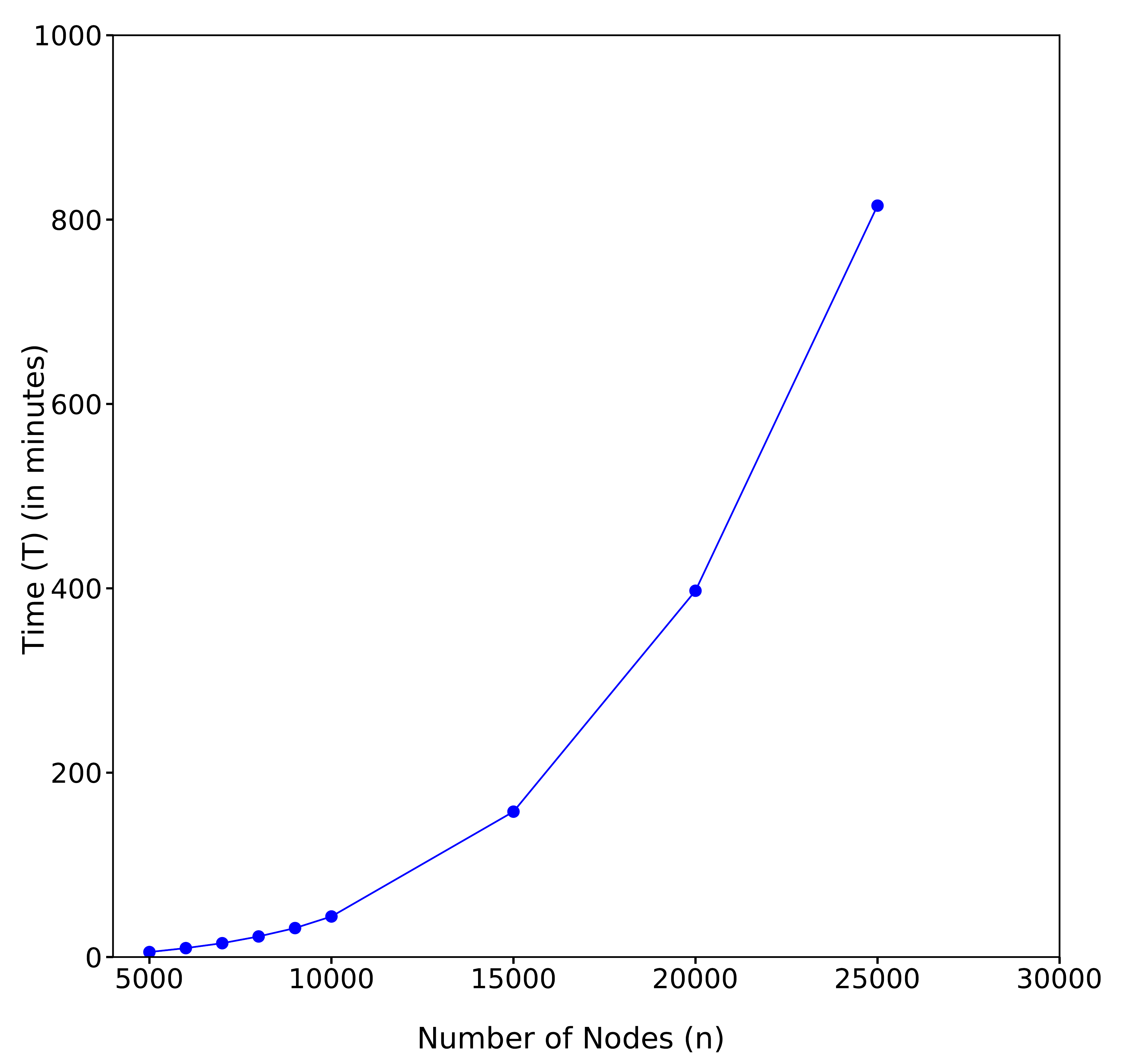}
\caption{\textbf{Computation Time}. The time taken by the network to completely loose its functionality was measured for network of different sizes. The figure shows the computational time for Barabasi-Albert (BA) networks of different sizes. It was observed that when the BA networks were attacked randomly, the time taken increased exponentially for networks with higher number of nodes. The other three attack models took very less computational time.  The decreasing order of time taken by the four attack models was observed to be Random attacks > Based on Probability > Targeted Attack - Model II > Targeted Attack - Model I. Similarly, an exponential behaviour was observed for ER networks of different sizes.} 
\end{figure*}

\begin{figure*}[htb]
\centering
\includegraphics[scale = 0.7]{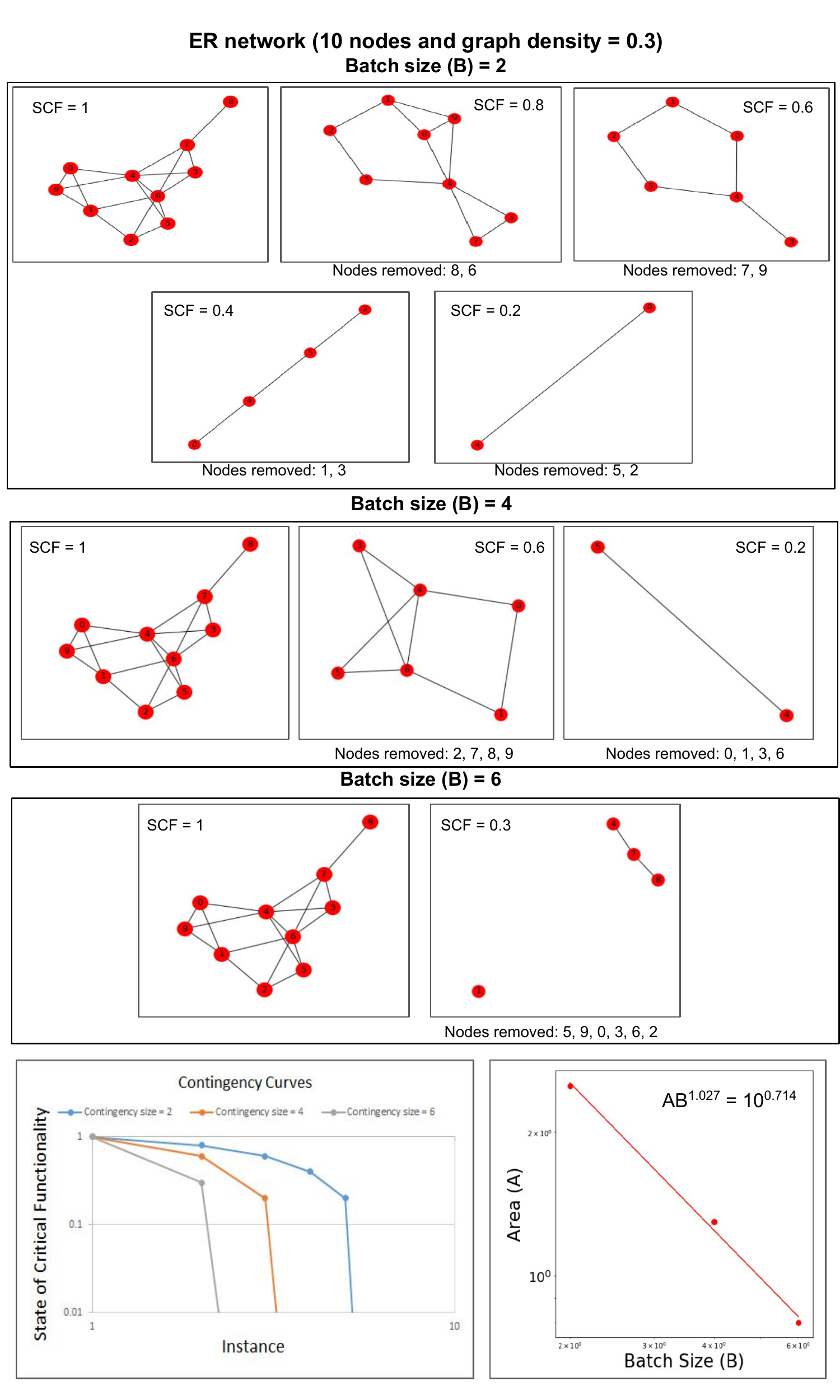}
\caption{\textbf{Toy Model}. The attack model for an ER network of 10 nodes, and graph density 0.3 has been depicted here. The graphs shown under the section of 'Batch size (B) = 2' has 5 sub figures. Each graph represents the state of network on removing 2 nodes. The last state of the network will have zero nodes present. Similarly, the state of network for attack sizes 4, and 6 has been shown. The contingency curve and a plot of Area (A) vs Batch size (B) has also been depicted in the figure. The best fit line in this case comes out to be $AB^{1.027}\ =\ 10^{0.714}$. The power of B in this equation exceeds '1'. If we consider batch sizes to be 2, 4, 6, 8, and 10, the power of B comes out to be 0.709. This is because of adding a few more points to the Area vs Batch size plot. The change in the best fit line of very small networks is significant as compared to large-scale networks.}
\end{figure*}

\begin{table}[tbhp]
\centering
\caption{Values of m in the equation $AB^m\ =\ constant$. It can be clearly seen that the value of m approaches 1 as the network size increases.}
\begin{tabular}{|l|r|r|r|r|}
\hline
\multirow{2}{*}{Network sizes} & \multicolumn{4}{c|}{Attack Models}                                                                           \\ \cline{2-5} 
                               & \multicolumn{1}{l|}{RAM} & \multicolumn{1}{l|}{TAM-I} & \multicolumn{1}{l|}{TAM-II} & \multicolumn{1}{l|}{TAM-III} \\ \hline
ER 400 nodes  & 0.917    & 0.974    & 0.886     & 0.965 \\ \hline
ER 800 nodes  & 0.970    & 0.998    & 0.983     & 0.993 \\ \hline
ER 1000 nodes                  & 0.981                    & 0.999                      & 0.972                       & 0.986                        \\ \hline
ER 2000 nodes                  & 0.996                    & 0.999                      & 0.995                       & 0.996                        \\ \hline
ER 3000 nodes                  & 0.998                    & 1                      & 0.997                       & 0.999                     \\ \hline
ER 5000 nodes                  & 0.999                    & 1                          & 0.999                       & 1                            \\ \hline
ER 10000 nodes & 1  & 1   & 1& 1           \\ \hline
ER 20000 nodes & 1  & 1   & 1& 1           \\ \hline
\end{tabular}
\end{table}

\end{document}